\documentclass[pra,twocolumn,floatfix,superscriptaddress,notitlepage,longbibliography]{revtex4-1}

\usepackage[normalem]{ulem}
\usepackage{graphicx, color, graphpap}
\usepackage{enumitem}
\usepackage{amssymb}
\usepackage{verbatim}
\usepackage{amsthm}
\usepackage{multirow}
\usepackage[colorlinks=true,citecolor=magenta,linkcolor=magenta]{hyperref}
\usepackage{centernot}
\usepackage[T1]{fontenc}
\usepackage{verbatim}
\usepackage{mathtools}
\usepackage{titlesec}
\usepackage{float}
\usepackage{caption}
\usepackage{subcaption}

\newcommand{\braket}[2]{\langle #1 \hspace{1pt} | \hspace{1pt} #2 \rangle}

\newcommand{\bramatket}[3]{\langle #1 \hspace{1pt} | #2 | \hspace{1pt} #3 \rangle}

\newcommand{\ket}[1]{|#1\rangle}              
              
\newcommand{\bra}[1]{\langle #1|}

\newcommand{\Tr}{{\rm Tr}}

\renewcommand{\geq}{\geqslant}
\renewcommand{\leq}{\leqslant}

\renewcommand{\vec}[1]{\boldsymbol{#1}}

\begin{document}
\title{Opportunities and Limitations in Broadband Sensing}

\author{Anthony M. Polloreno}
\thanks{ampolloreno@gmail.com}
\author{Jacob L. Beckey}
\affiliation{JILA, NIST and University of Colorado, Boulder, Colorado 80309, USA}

\author{Joshua Levin}
\affiliation{JILA, NIST and University of Colorado, Boulder, Colorado 80309, USA}
\affiliation{Quantinuum, 303 S. Technology Ct., Broomfield, Colorado 80021, USA}

\author{Ariel Shlosberg}
\author{James K. Thompson}
\affiliation{JILA, NIST and University of Colorado, Boulder, Colorado 80309, USA}

\author{Michael Foss-Feig}
\author{David Hayes}
\affiliation{Quantinuum, 303 S. Technology Ct., Broomfield, Colorado 80021, USA}

\author{Graeme Smith}
\affiliation{JILA, NIST and University of Colorado, Boulder, Colorado 80309, USA}

\begin{abstract} 
Detecting a signal at an unknown frequency is a common task, arising in settings from dark matter detection to magnetometry. For any detection protocol, the precision achieved depends on the signal’s frequency and can be quantified by the quantum Fisher information (QFI). To study limitations in broadband sensing, we introduce the integrated quantum Fisher information and derive inequality bounds that embody fundamental trade-offs in any sensing protocol. Our inequalities show that sensitivity in one frequency range must come at the cost of reduced sensitivity elsewhere. For many protocols, including those with small phase accumulation and those consisting of $\pi$-pulses, we find the integrated quantum Fisher information scales linearly with $T$. We also find protocols with substantial phase accumulation that can have integrated QFI that grows quadratically with $T$, and prove that this scaling is asymptotically optimal. These protocols may allow the very rapid detection of a signal with unknown frequency over a very wide bandwidth. We discuss the implications of these results for a wide variety of contexts, including dark matter searches and dynamical decoupling. Thus we establish fundamental limitations on the broadband detection of signals and highlight their consequences.

\end{abstract}
\maketitle

\section{Introduction}Quantum systems, from atoms to SQUIDs, can be excellent sensors. 
Indeed detecting weak signals requires the consideration of quantum effects. Even better, entangled detectors are well-known to be more sensitive than their unentangled counterparts \cite{Giovannetti2004}. In practice, quantum sensors have been used for dark matter searches, entanglement-enhanced magnetometry, microwave clocks, and matterwave interferometers. \cite{Backets2021DarkMatterLehnert, PolzikSqueezedMagnetometer2010, VladanEntangledClockLifetime2010, KasevichSqueezedClock2020, greve2021entanglementenhanced, WinelandSixAtomCatState2005, BlattGHZState14Qubit2011, lawrie2019quantum, pooser2020truncated, PhysRevD.23.1693}.

 The quantum Fisher information (QFI) captures the performance of a parameter estimation protocol \cite{helstrom1967minimum}. For a pure state parameterized by $\theta$, $\rho_\theta := \ket{\psi_\theta}\bra{\psi_\theta}$, the QFI is 
\begin{equation}
    J(\theta) = 2\text{Tr} \left[(\partial_\theta\rho_\theta)^2\right],
\end{equation}
\cite{Fujiwara1995}. The QFI tells us, via the Cramer-Rao bound \cite{braunstein1994statistical}, how well an unbiased estimator of $\theta $ can approximate its true value. In particular, given $m$ copies of $\ket{\psi_\theta}$ the variance of any unbiased estimator $\hat{\theta}$ must satisfy $\text{Var}(\hat{\theta}) \geq \frac{1}{mJ(\theta)}$.  This bound can be saturated,
so the more quantum Fisher information a protocol has, the better one can estimate $\theta$ \cite{braunstein1994statistical}.

We consider a coupling between the qubit and the signal as 
\begin{align}\label{Eq:SignalHamiltonian}
H(t) = \mu B\cos(\omega t + \varphi) Z,    
\end{align}
where $\mu$ is the magnetic moment of the qubit.
For instance, we may wish to estimate the strength of an 
AC magnetic field \cite{weiner2012,fiderer2019maximal, clarke2004alex, budker2007optical, dang2010ultrahigh, degen2008scanning,arcizet2011single, kolkowitz2012coherent, bal2012ultrasensitive}. While we have included $\varphi$ in the analysis in Eq.~\eqref{eq:with_phi}, we have omitted it for convenience elsewhere.

To gather information about this Hamiltonian, we need to establish a protocol, which generically consists of preparing the sensing qubit in an initial state, applying a time-dependent control sequence, and finally performing a measurement. The performance of a protocol will depend on the frequency of the signal $\omega$. For example, preparing $\ket{+}$ followed by free evolution for time $T$ and measurement in the $\ket{\pm}$ basis is optimal for $\omega =0$ but performs poorly for $\omega\gg 1/T$. In fact, we will show that trade-offs in sensitivity at different frequencies are inevitable. We make this quantitative by considering the integrated QFI (IQFI).

While this may be different than the integrated sensitivity of a single protocol, it provides an analytically tractable method of analysis which is tight in many cases of interest, namely weak fields and protocols consisting only of $\pi-$pulses. In general, a choice of protocol includes a choice of measurement basis, as defined above. In the majority of examples we consider in this paper, the optimal measurement basis is frequency independent, so that the the optimization over the final measurement in the definition of QFI can be ignored. By bounding this integral we formalize the idea that there is a fundamental tension between having sensitivity in different frequency bands. 

The longer we observe a signal, the more we can expect to learn about it. Thus, it is no surprise that the IQFI will typically grow with the duration $T$ of a protocol.  In fact, we find a number of constraints on how IQFI grows with $T$. First, we find that any protocol starting on the equator of the Bloch sphere that involves only $\pi$-pulses has an IQFI of $2\pi\zeta^2 T$, where $\zeta\equiv \mu/\hbar$ is the inverse gyromagnetic ratio of the system being used as a sensor \cite{phillips1977magnetic}. Second, for an arbitrary protocol with $\zeta BT \ll 1$, the IQFI is close to $2\pi\zeta^2 T$. Then we study a particular protocol that significantly exceeds $2\pi\zeta^2 T$ --- by continuously driving a spin with a transverse $gX$ term, our protocol has a peak sensitivity around $2g$, with IQFI scaling quadratically with time. We further show that the IQFI can not exceed quadratic scaling with time, so that this protocol is in a sense optimal. However, the practical restriction to the small signal regime $\zeta BT\ll 1$ is not uncommon, due in part to phase ambiguities that may arise in the accrued phase if one begins to leave this regime.

Beyond signal estimation and detection, our results can be applied to better understand the performance of dynamical decoupling \cite{viola1999dynamical, viola2003robust, khodjasteh2009dynamical, santos2006enhanced}. Here we find that for many dynamical decoupling protocols, the average (over the initial state) IQFI is at least linearly proportional to $T$. As a result, dynamical decoupling can at best move the noise sensitivity of a qubit around in frequency space, rather than eliminating sensitivity at all frequencies. This is reminiscent of the filter functions discussed in \cite{Bylander2011} which considers CPMG sequences \cite{meiboom1958modified, carr1954effects, PhysRevB.77.174509}, a subset of the protocols considered in this work.

We note that the setting under consideration is different from the waveform estimation studied in \cite{Tsang2011}. That work studied how to simultaneously estimate a large number of parameters representing the full time series of a waveform. We consider the sensing problem of estimating a single Fourier amplitude, i.e. the systems we consider couple to a monochromatic signal $B\cos{(\omega t)}$. The relative simplicity of this setting admits a global analysis of the performance of an arbitrary protocol at different frequencies. 

\section{Preliminaries} We consider Hamiltonians of the form Eq.~(\ref{Eq:SignalHamiltonian}),
motivated by a spin-$\frac{1}{2}$ particle in a magnetic field. We first consider estimation protocols composed of instantaneous, arbitrary unitary rotations $P_i$ followed by periods of free evolution. The choice of measurement at the end of the protocol is assumed to be optimal in the sense that it maximizes the Fisher information of the resulting classical probability distribution. For a state starting in the $+1$ $X$ eigenstate, we have the final state
\begin{equation}\label{eq:discrete_evolution}
    \ket{\psi(T, \omega, B)} = U_{N-1}(t_N, t_{N-1})P_{N-1}...U_0(t_1, t_0)P_0\ket{+},
\end{equation}
where $t_N = T$ and $U(t_{i+1}, t_i)$ is the time evolution operator under the Hamiltonian in Eq.~\eqref{Eq:SignalHamiltonian} between times $t_i$ and $t_{i+1}$.

Given Eq.~\eqref{eq:discrete_evolution}, the QFI tells us
how well we can estimate $B$. We write $J(B|\omega)$ to indicate that Fisher information with respect to $B$ will in fact depend on the signal frequency $\omega$. Writing $\ket{\phi} := \partial_B\ket{\psi(T, \omega, B)}$ and $\ket{\psi} := \ket{\psi(T, \omega, B)}$ the QFI can be expressed as \cite{Fujiwara1995,Pang2017}
\begin{equation}
\label{eq:pure_qfi}
J(B|\omega) = 4(\braket{\phi}{\phi} + \text{Re}\{\braket{\phi}{\psi}^2\}).
\end{equation}
To understand the total sensitivity of a protocol across all frequencies we define the integrated QFI (IQFI) for a protocol with total evolution time $T$ as \footnote{Note that the integral in Eq.~\eqref{eq:IQFI} always exists since the QFI is bounded and asymptotically we have 
$J(\rho_\omega) \sim (\sum^{N-1}_{i=0} \Theta_i)^2 \sim 1/\omega^2$}

\begin{equation}\label{eq:IQFI}
    K(T) = \int_0^{\infty}d\omega  J(B|\omega).
\end{equation}

Statistically, one way of understanding the IQFI is as an unnormalized expected quantum Fisher information (EQFI) for estimating a magnetic field amplitude at an unknown frequency with a uniform prior. While there is no uniform distribution over the positive real line, because the QFI vanishes at high frequencies, it is possible, in principle, to establish a cutoff at which the IQFI is approximately an unnormalized expected quantum Fisher information. As we show in Appendix~\ref{sec:gen_proof}, this cutoff is independent of time, and so can be varied independently of protocol duration. While the EQFI is an interesting statistical quantity, because this equivalence is only approximate, and because we are interested in a measure of how the total (unnormalized) bandwidth sensitivity of a protocol changes with time, we consider the IQFI rather than the EQFI. One way of understanding this distinction is that the former has units of time, while the latter has units of time squared, and so they measure different physical quantities.

Nevertheless, it is possible to give an interpretation of our results in the Bayesian setting. For the choice of a delta function prior, it is known that EQFI grows at most quadratically with time \cite{Pang2017}. In this work, however, we are interested in understanding the optimal performance of a protocol given an arbitrary prior. We will show that the optimal performance for the uniform prior is also quadratic in time, and thus for any prior quadratic scaling is also optimal. 

To see this, consider the following integral where $\mu$ is the uniform prior, $\xi$ is any other prior, and $\Omega$ is some cutoff frequency beyond which the QFI and $\xi$ are both negligible,
\begin{equation}
    \int_0^{\Omega} d\xi(\omega) J(B|\omega) \leq M\int_0^{\Omega} d\mu(\omega) J(B|\omega).
\end{equation}
In this equation $M = {\rm max}(\xi(\omega))/\mu_0$ where $\mu_0$ is the value of the uniform prior. Any error in this expression is independent of time and can be suppressed by increasing $\Omega$, and so we see that the asymptotic behavior of the EQFI with a uniform prior provides an upper bound on the EQFI with any prior.


\section{Ramsey and $\pi$-pulse protocols}Within the family of control sequences consisting of instantaneous rotations interleaved with free-evolution, we now consider Ramsey spectroscopy, where a qubit is prepared on the equator of the Bloch sphere and allowed to freely precess. Defining $\Theta(t_{i+1}, t_i) = [\sin{(\omega t_{i+1}} - \sin{(\omega t_{i}}]/(\hbar\omega)$, we find $\ket{\phi} = -i\zeta\Theta(T, 0)\ket{\psi}$
and Eq.~\eqref{eq:pure_qfi} gives $J(B|\omega) =  4\zeta^2\Theta^2(T, 0)$. Defining $\zeta\equiv \mu/\hbar$, the IQFI follows as
\begin{align}\label{eq:with_phi}
    K(T) &= 4\zeta^2\int_0^{\infty} d\omega \frac{(\sin{(\omega T+\varphi)}-\sin{(\varphi)})^2}{\omega^2},\\
    &= 2\zeta^2 T(\pi-\ln(4)\sin{(2\varphi)}).
\end{align}
If $\varphi$ is unknown and therefore random in each experiment, averaging over $\varphi$ gives $K(T)=2\pi\zeta^2T$, but if $\varphi$ is known and we wish to maximize the IQFI, we would lock the experimental sequence to $\varphi=3\pi /4$ to get  $K(T)=2\zeta^2T(\pi+\ln{4})$. For convenience, in what follows we assume $\varphi = 0$.

Now consider a protocol applying $\pi$-pulses at times $t_1$,\dots, $t_{N}=T$. At time $t_i$ we apply either $X$, $Y$ or $Z$. Additionally, we can apply any unitary that leaves the expectation value of $Z$ invariant \footnote{Consider a $\pi-$rotation about the axis halfway between the X- and Y- axes, given by $\exp{(-i\frac{\pi}{2\sqrt{2}}(X+Y)}$)}. Then we have 

\begin{eqnarray}\label{eq:pipulse}
\begin{aligned}
    K(T) &= 4\zeta^2\int_0^{\infty}d\omega(\sum_{i=0}^{N-1}\Theta(t_{i+1}, t_i))^2\\
     &= 2\pi\zeta^2 T.
      \end{aligned}
\end{eqnarray}

$K(T)$ was derived assuming that the system was initialized in the $\ket{+}$ state. If we instead integrate over all initial states we find that the average IQFI is given as $\frac{4\pi\zeta^2 T}{3}$.
This average has important implications for dynamical decoupling protocols based on $\pi$-pulses.  It shows that for such protocols a qubit must maintain a sensitivity to environmental noise over a substantial frequency range. Indeed, in the presence of white noise any $\pi$-pulse protocol will leave the qubit equally degraded by the noise. Specifically, we imagine that a qubit is subjected to a background power spectrum of magnetic field noise fluctuations whose noise spectrum is flat. A lower bound on the IQFI implies a lower bound on how much of this noise spectrum the qubit will experience, and therefore on its decoherence. For noise with more structure, these protocols do not allow sensitivity to noise to be eliminated, but can simply move that sensitivity to a frequency range where the environmental noise is fairly low. 

\section{B=0 Bound} We now present an argument to bound $K(T)$ at $B=0$, and approximately bound $K(T)$ for short times and weak magnetic fields. Consider protocols with initial state $\ket{\psi_0}$ and a control Hamiltonian $H_0(t)$ in addition to the signal Hamiltonian $H(t)$. We can then write $\ket{\psi(T, \omega, B)} = U\mathcal{U}\ket{\psi_0}$, where $U$ is the time evolution due to $H_0$ and $\mathcal{U}$ is the interaction picture time-evolution, given by
\begin{align}
    \mathcal{U} = 1 - i\zeta B\int_0^T\cos(\omega \tau)Z_I(\tau)d\tau + O(B^2), 
\end{align}
where we have used $Z_I(\tau)$ to express $Z$ in the interaction picture.
Then we have
\begin{align}
    \partial_B\ket{\psi(T, \omega, B)}\Big|_{B=0} = -i\zeta U\int_0^T\cos(\omega \tau)Z_I(\tau)d\tau\ket{\psi_0}.
\end{align}
Substituting into Eq.~\eqref{eq:pure_qfi} and integrating, we find that the IQFI is at most $2\pi \zeta^2 T$. For small $B$ and $T$, the next term on dimensional grounds should be $\mathcal{O}(B^2T^3)$, since we can show the term linear in $B$ is zero. This dimensional analysis assumes that there are no other dimensionful quantities - for instance, if the interpulse spacings are not functions of $T$, then there may be other terms. Thus we find 
\begin{equation}
    K(T) \leq 2 \pi\zeta^2 T + O(\zeta^4B^2T^3)
\end{equation}
The full proof is provided in Appendix~\ref{app:pert}. This shows that for small magnetic fields and short times we should expect a roughly linear scaling of the IQFI.

\section{Entangled probe advantage}
From standard results \cite{holland1993interferometric, bollinger1996optimal} we expect that entangled inputs can outperform this bound. Indeed an $n-$qubit GHZ state
\begin{equation}
    \ket{GHZ}_n = \frac{1}{\sqrt{2}}(\ket{0}^{\otimes n} + \ket{1}^{\otimes n}).
\end{equation}
accumulates phase $n$ times more rapidly, so that an analogous argument gives an IQFI at $B=0$ of 
\begin{equation}\label{eq:entangled_discrete}
    K(T) = 2\pi n^2\zeta^2 T.
\end{equation}
Conversely, product input states can be reduced to the single qubit example, since $J(\rho^{\otimes n}) = nJ(\rho)$, so that for $n$ qubits starting in a product state, again with $B=0$, we have
\begin{equation}\label{eq:separable_discrete}
    K(T) \leq 2\pi n\zeta^2 T.
\end{equation}
 So, while entanglement allows us to increase the coefficient in front of $T$, IQFI still increases linearly with time.

\section{Quadratic scaling of IQFI}
We have seen that $\pi$-pulse protocols and protocols with $BT\zeta \ll 1$ have IQFI that scales no faster than $2\pi\zeta^2 T$. Even an entanglement-enhanced protocol gives linear scaling of IQFI with $T$, albeit with an improved coefficient. We now study a simple protocol with IQFI scaling quadratically in $T$.  

We consider a continuous-time protocol which applies a transverse field, $\hbar g X$, to the sensing qubit. This gives a full Hamiltonian of 
\begin{align}\label{eq:gX}
H(T) = \hbar g X + \mu B \cos{(\omega T)} Z.
\end{align}
Assuming $\omega \sim 2g$, $BT\zeta \gg 1$ and using the rotating wave approximation \cite{allen1987optical}, we find
\begin{align}
J(B|\omega) &\sim  \frac{(\zeta T)^2}{(1+(\frac{\omega-2g}{B \zeta})^2)^2} 
\end{align}
which we can integrate from $\omega = g$ to $\omega = 3g$ to get a lower bound of
\begin{align}
    K(T) \gtrsim \zeta^2T^2 \left(\frac{ g }{1+ \frac{g^2}{\zeta^2 B^2}}+\zeta B\tan ^{-1}\left(\frac{g}{\zeta B}\right)\right).
\end{align}
A detailed derivation is given in Appendix \ref{sec:RWA}. Representative dynamics in this regime are shown in Fig. 3 and 4 of \cite{PhysRevA.86.032313}.

\begin{figure}
    \includegraphics[width=.45\textwidth]{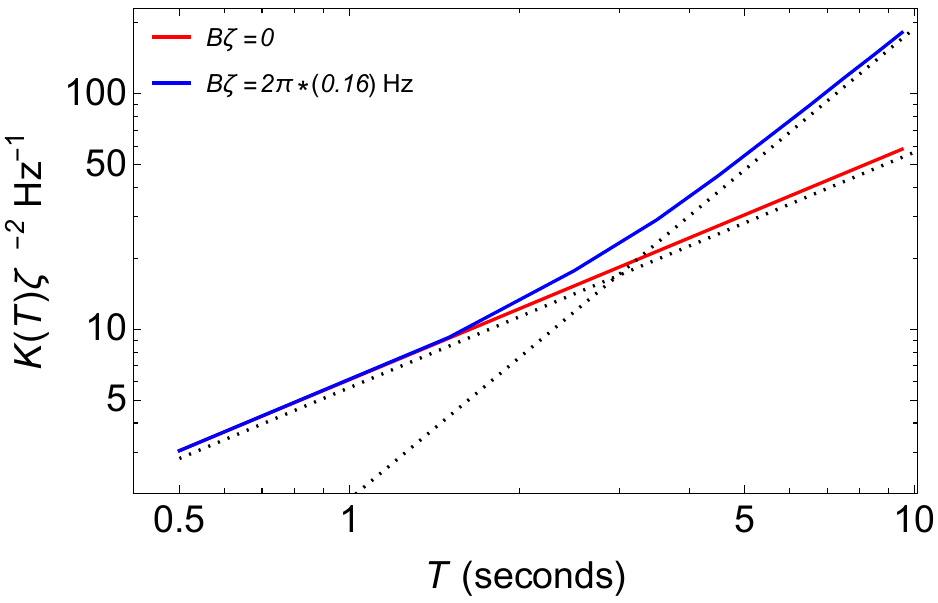}
    \caption{IQFI ($K(T)$) for using $X_{\pi/2}$ pulses. Depending on the magnetic field strength we observe a more rapid accumulation of IQFI. The dashed lines are $\sim T$ and $\sim T^2$ scalings to guide the eye. For non-zero magnetic field, we see that a crossover from quadratic to linear scaling occurs when when $BT\zeta\ll 1$ (where the perturbative results are valid).}
    \label{fig:discrete}
\end{figure}

\begin{figure}
  \begin{subfigure}[t]{0.22\textwidth}

    \includegraphics[scale=.28]{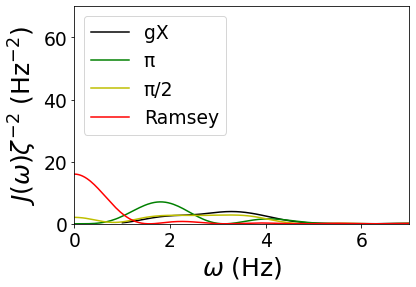}
    \caption{\textbf{}}
    \end{subfigure}
    \begin{subfigure}[t]{0.22\textwidth}

    \includegraphics[scale=.28]{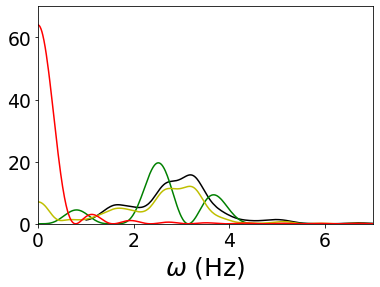}
    \caption{\textbf{}}
    \end{subfigure}
    \begin{subfigure}[t]{0.22\textwidth}

    \includegraphics[scale=.28]{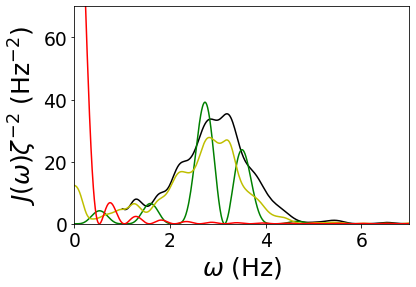}
    \caption{}
    \end{subfigure}
    \begin{subfigure}[t]{0.22\textwidth}

    \includegraphics[scale=.28]{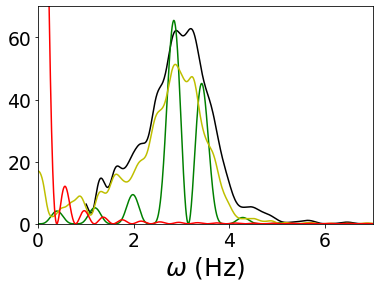}
    \caption{}
    \end{subfigure}
    
    \caption{QFI as a function of frequency for some protocols considered in this paper, for different protocol times $T$ [(a) $T=2$,  (b) $T=4$, (c) $T=6$, and  (d) $T=8$ seconds], with $B\zeta = 1$ Hz. Ramsey has a large DC QFI, as expected, and very little ability to detect any AC signal. $\pi$-pulses, on the other hand, can be used to measure signals at higher frequencies, as might be expected from spectroscopy techniques such as CPMG. The gX protocol, with $g=2\pi\times \frac{1}{4}$Hz in this example, is seen to be sensitive near $2g$. Moreover, it is seen to be sensitive over a broad bandwidth. The $\pi$-protocol shown here consists of $\pi$ rotations about the $X$ axis, at each integer value of time. The $\pi/2$-protocol consists of $\pi/2$ rotations about the $X$ axis, every half second. Similar to the $\pi$-protocol, the $\pi/2$-protocol has AC sensitivity, as well as more broadband sensitivity.}
    \label{fig:qfivomega}
\end{figure}

In Fig.~\ref{fig:qfivomega} we numerically study this protocol. We first approximate it by a discrete pulse-based protocol, described by Eq.~\eqref{eq:discrete_evolution} and given by Trotterization, that intersperses instantaneous rotations around the $X$-axis by periods of free evolution under the magnetic field. For time $T$, we can approximate evolution under Eq.~\eqref{eq:gX} by $m$ periods of free evolution of duration $\frac{T}{m}$ separated by rotations of angle $\frac{T\pi}{m}$ about the $X$-axis. Indeed, in Fig.~\ref{fig:discrete}, we see quadratic scaling choosing $m=2T$. In Fig.~\ref{fig:qfivomega} we compare the QFI of the $m=T$ and $m=2T$ cases with the $gX$ protocol and a Ramsey protocol, where it is evident that both the $m=2T$ protocol and the $gX$ protocol accumulate IQFI more rapidly than the other two protocols. These discrete protocols with quadratic scaling of IQFI have the property that the number of pulses scales with the total time of the protocol.  In fact, this is necessary, as we will now see.

Consider a protocol with $N$ pulses $P_i$ applied between periods of free evolution $U_i$. 
Because $\braket{\psi}{\psi}=1$, $\partial_B\braket{\psi}{\psi} = 0$, we have $\braket{\phi}{\psi} = - \braket{\psi}{\phi}$. This means that $Re(\braket{\phi}{\psi}) = 0$, and so $\braket{\psi}{\phi}^2$ has real part that is non-positive. Thus from Eq.~\eqref{eq:pure_qfi} we see
\begin{equation}
\label{eq:pure_qfi_bound}
J(B|\omega) \leq 4\braket{\phi}{\phi}.
\end{equation}

We see that $\braket{\phi}{\phi} = \zeta^2 \sum_{i,j}^N \Theta_i\Theta_jV_{ij} = \zeta^2\vec{\Theta}^{T}V\vec{\Theta}$, where $\vec{\Theta}$ is a vector whose $i^{th}$ entry is $\Theta_i\equiv \Theta(t_{i+1}, t_i)$.  $V$ is an $N\times N$ complex matrix with entries of norm at most 1, so its eigenvalues have norm at most $N$. We thus find
\begin{eqnarray}\label{eq:discrete-bound}
&K(T) = \int_0^{\infty}d\omega J(\omega) \leq 4\int_0^{\infty} d\omega \braket{\phi}{\phi} \\&=4\zeta^2\int_0^{\infty}d\omega \vec{\Theta}^{T}V\vec{\Theta} \leq 4\zeta^2\int_0^{\infty}d\omega N |\vec{\Theta}|^2 = 2\pi N\zeta^2 T.\nonumber
\end{eqnarray}
We will now see that the bound Eq.~(\ref{eq:discrete-bound}) can be used to show that the IQFI can scale at most quadratically with time, so that up to a constant the IQFI of the $gX$ protocol grows as rapidly as possible in time. 

A continuous-time protocol involves a control Hamiltonian $G$ that gives a total Hamiltonian of the system
\begin{equation}\label{eq:continuous-sensing}
H(t) = \hbar G(t) + \mu B\cos(\omega t) Z.    
\end{equation}

We can Trotterize this evolution into a discrete sequence \cite{poulin2011quantum} like those considered in proving Eq.~(\ref{eq:discrete-bound}). Then, by linearity of the derivative we can constrain the derivative of the Trotterized evolution to be close to the actual derivative. Using $U(t)$ to refer to the continuous time protocol, and $\hat{\epsilon}''(t) = \partial_BU(t) - \partial_BU'(t)$, we see from Eq.~(\ref{eq:pure_qfi_bound}),

\begin{align*}\label{eq:naive_inequality}
    J(\omega) \leq J_d(\omega) + 8Re(\bramatket{+}{\hat{\epsilon}''(T)^{\dagger}(\partial_B U'(T) + 4\hat{\epsilon}''(T))}{+}),
\end{align*}
where we have used $J_d(\omega)$ to denote the QFI of the protocol given by $U'(t)$. As we show in Appendix~\ref{app:gen_proof}, this gives a bound of
\begin{eqnarray}
    K(T) \leq \frac{2\pi\zeta^2 T^2}{\delta t} &+& \alpha(\delta t, \Omega, \mu B, \hbar||G||)T^2\nonumber,
\end{eqnarray}
where $||G||$ is maximum spectral norm of the control, ${\rm max}_{t\leq T}||G(t)||$ and $\alpha$ is a function that controls the error in approximating the IQFI of the continuous protocol by the IQFI of the Trotterization. Crucially it does not depend on $T$. This proves that $K(T)$ scales at most quadratically in time. Thus, our examples with quadratic scaling are asymptotically optimal in the amount of IQFI they accumulate. This bound can be extended to consider both entangled probes and separable probes as shown in Appendix~\ref{sec:gen_proof}, and thus establishes a fundamental limit on the broadband sensitivity of detectors.

\section{Conclusions}
The QFI provides an ultimate bound on how well a quantity can be estimated, in our case the amplitude of a sinusoid with fixed frequency. Integrating the QFI over all frequencies, we found fundamental limits on the broadband performance of quantum sensors. For tasks such as axion detection \cite{budker2014proposal}, this implies that spectral sensitivity is a scarce resource that needs to be carefully considered when designing metrological protocols. While conventional spectroscopy protocols such as Ramsey interferometry and CPMG \cite{meiboom1958modified, carr1954effects} consist only of $\pi$-pulses, and therefore linearly accumulate IQFI, we found both continuous and discrete protocols that quadratically increase this accumulation. Moreover, we have shown that this is asymptotically the largest scaling one can achieve.

We have seen that there are protocols with IQFIs that scale as both $\mathcal{O}(\zeta^2 T)$ and $\mathcal{O}(\zeta^3 BT^2)$, but which is better?  It depends---if the goal is sensitivity to a wide range of frequencies, $\mathcal{O}(\zeta^3 BT^2)$ may allow the protocol to work over a wider frequency band. If the goal is sensitivity to a very narrow frequency range, $\mathcal{O}(\zeta^2 T)$ protocols may have support over a small band. Thus, we may see enhancements when searching for a weak signal over a wide frequency range. In such a setting, long integration times could give a quadratic enhancement of the accumulated QFI compared to the $\mathcal{O}(\zeta^2 T)$ protocols.

Additionally if $T_d$ is a characteristic decoherence time of the system, then we are practically constrained to $T < T_d$. Thus for $B \ll 1/(\zeta T_d)$ and fixed $T_d$, we also find $\zeta BT\ll1$. In these contexts, our results show that sensing should be expected to be limited by the bound of $2\pi\zeta^2 T$. Physically, this corresponds to when the peak angular excursion of the Bloch vector is much less than $\pi$. For weak fields, if we can sense for long times so that $\zeta B T \centernot\ll 1$, then we can accumulate quadratically more IQFI.

The $gX$ protocol is sensitive to frequencies around $2g$, making it an excellent candidate for broadband detection around a particular frequency. It is an open question how to design optimal metrological protocols with sensitivity spread evenly over wide bands. Techniques like GRAPE \cite{Khaneja2005} may be useful for this task \cite{Liu2017}.
 
Many dynamical decoupling protocols consist solely of $\pi$-pulses (e.g., \cite{UDD98}). Such techniques may be described by Eq.~\eqref{eq:discrete_evolution}. Consequently, our results show these decoupling strategies are fundamentally limited - while they can move noise sensitivity, they cannot remove it. We leave open whether such bounds apply to arbitrary protocols. 

Our key conceptual contribution is the idea that IQFI can be used to understand the trade-offs inherent in broadband sensing. In some cases, this metric provides a conservation law that can be summarized by the slogan ``no free QFI''. In particular, in the case where the interaction picture operator being sensed ($Z_I(t)$) commutes with itself at all times and in the small signal limit ($\zeta BT\ll 1$) we showed that QFI at one frequency ultimately comes at the cost of less QFI at another frequency. This is also true for sequences consisting only of $\pi$ pulses, when the sensor state begins on the equator. Moreover we have shown that for \textit{any} protocol there is a limited amount of IQFI that can be accumulated. This demonstrates that while broadband sensing is possible, there is an upper limit on how wide the bandwidth of a given protocol can be if one desires a certain sensitivity. We do not currently know if other classes of control protocols yield strict conservation laws, and we leave this to future work. 

The bounds on IQFI that we have found on single qubit initial states can be extended, via Eq.~(\ref{eq:separable_discrete}), to arbitrary separable states due to the additivity and convexity of the QFI \cite{feng2017quantifying, toth2014quantum}. As Eq.~(\ref{eq:separable_discrete}) applies only to separable probe states, we can think of it as a kind of standard quantum limit that cannot be exceeded without entanglement. Indeed, we see that an n-qubit cat state can significantly exceed the $2\pi n\zeta^2 T$ performance of unentangled $\pi$-pulse-based protocols. This points to the possibility of using IQFI as a form of entanglement witness, so that the quantum Fisher information at any particular frequency may be consistent with a separable state but the breadth of such sensitivity can only be explained by an entangled state. In general, however, we find that the quadratic-in-time asymptotic upperbound of the IQFI holds for both separable and entangled probes. Finally, another interesting open question is if other transformations of QFI spectra might generate new insights into broadband sensing limitations.

\section{Acknowledgments}
We thank Charles Marrder for reading and commenting on our manuscript. JLB thanks Simon J{\"a}ger for helpful discussions. DH acknowledges helpful discussions with Carl Caves. This work was partially supported by the NSF JILA PFC grant 1734006. AMP acknowledges funding from a NASA Space Technology Graduate Research Opportunity award.

\onecolumngrid
\section{Appendix}
\label{sec:appendix}

\subsection{Derivation of Average IQFI for Instantaneous $\pi$-Pulses}
\label{Derivation of Average IQFI for Instantaneous Pi-Pulses}
Using the identity
\begin{equation}
    [A, UBU^{\dagger}] = U[U^{\dagger}AU, B]U^{\dagger}
\end{equation}
we find the N-pulse QFI is given by

\begin{gather*}
    J(\omega) = -2 \Tr\Big(\Big(\left[U_0^{\dagger}P_0^{\dagger}U_1^{\dagger}P_1^{\dagger}...U_{N-1}^{\dagger}P_{N-1}^{\dagger}ZP_{N-1}U_{N-1}...P_0U_0,\frac{X}{2}\right]\Theta_{N-1}+...+\left[Z,\frac{X}{2}\right]\Theta_0\Big)^2\Big) 
\end{gather*}

Integrating over all frequency and using the orthogonality condition
\begin{equation}
    \int d\omega \Theta_i\Theta_j = 2\pi(t_{i+1}-t_i)\delta_{ij},
\end{equation}
we see that when this square is integrated this will sum to be $2\pi T$, so that for a protocol starting in the $\ket{+}$ state and using only $\pi$-pulses we find 
\begin{eqnarray}
    K(T) = 2\pi T.
\end{eqnarray}

We can also average over the input state, denoted with angle brackets $\langle \cdot \rangle$, to find
\begin{align}
    \langle K(T) \rangle &= \frac{1}{4\pi}\int_{0}^{2\pi} d\beta \int_{0}^{\pi} d\alpha \sin{(\alpha)} 2\pi \zeta^2T\sin^2{\alpha}, \label{eq:avgIQFIpi}
\end{align}
where one copy of $\sin$ comes from the integration measure, and two copies come from the off-diagonal element of the density matrix. This gives the result from the main text, 
\begin{align}
    \langle K(T) \rangle &= \frac{4 \zeta^2 \pi}{3}  T.
\end{align}
\subsection{Perturbative Expansion to $\mathcal{O} (B^2)$}\label{app:pert}
Recall that the pure state QFI can be expressed as 
\begin{align}
    J(\rho) = 4\braket{\phi}{\phi} + 4 \text{Re}\{\braket{\phi}{\psi}^2\}.
\end{align}
As shown in a previous note, the second term above is in general non-positive and thus if all we seek is an upper bound on the QFI, we can simply consider the first term. We can write the time evolution operator as 
\begin{align}
    \mathcal{U} &= 1 - iB\int_0^t\cos(\omega \tau)Z_I(\tau)d\tau - B^2 \int_0^t d\tau_2 \int_0^{\tau_2} d\tau_1 \cos(\omega \tau_2)\cos(\omega \tau_1)Z_I(\tau_2) Z_I(\tau_1)\\
    &+  iB^3 \int_0^t d\tau_3 \int_0^{\tau_3} d\tau_2 \int_0^{\tau_2} d\tau_1 \cos(\omega \tau_3) \cos(\omega \tau_2)\cos(\omega \tau_1)Z_I(\tau_3)Z_I(\tau_2) Z_I(\tau_1) + \mathcal{O}(B^4),
\end{align}
where we have used $Z_I(t)$ to express $Z$ in the interaction picture. We expand to order $B^3$ because, when differentiated with respect to $B$, this yields a term proportional to $B^2$. To simplify notation, let 
\begin{align}
    \mathcal{U} &:= 1 - iB I_1 - B^2 I_2 + iB^3 I_3 + \mathcal{O}(B^4).
\end{align}
Then, the time-evolved quantum state will be 
\begin{align}
    \ket{\psi}:= \ket{\psi(t)} &= U_0 \mathcal{U} \ket{\psi (0)},
\end{align}
and the derivative of this state is then
\begin{align}
    \ket{\phi} &:= \partial_B \ket{\psi},\\
    &=\partial_B \ket{\psi(t)},\\
    &= \partial_B (U_0 \mathcal{U} \ket{\psi (0)}),\\
    &= (-i U_0 I_1 -2B U_0 I_2 +i3B^2 U_0 I_3)\ket{\psi(0)} + \mathcal{O}(B^3).
\end{align}
The inner product of this vector with itself is,
\begin{align}
    \braket{\phi}{\phi} &= \bra{\psi(0)} (+i I_1^{\dagger} U_0^{\dagger} -2B I_2^{\dagger} U_0^{\dagger} -i 3B^2 I_3^{\dagger} U_0^{\dagger})(-i U_0 I_1 -2B U_0 I_2 +i3B^2 U_0 I_3)\ket{\psi(0)} + \mathcal{O}(B^3),\\
    &= \bra{\psi(0)} I_1^{\dagger} I_1 \ket{\psi(0)} - i2B \bra{\psi(0)} I_1^{\dagger} I_2 \ket{\psi(0)} - 3B^2 \bra{\psi(0)} I_1^{\dagger} I_3 \ket{\psi(0)} +\\
    &+ i2B \bra{\psi(0)} I_2^{\dagger} I_1 \ket{\psi(0)}+ 4B^2 \bra{\psi(0)} I_2^{\dagger} I_2 \ket{\psi(0)} -3B^2 \bra{\psi(0)} I_3^{\dagger} I_1 \ket{\psi(0)} +\mathcal{O}(B^3).\\
\end{align}
Now we assume, and verify later, that $ \bra{\psi(0)} I_1^{\dagger} I_2 \ket{\psi(0)} = \bra{\psi(0)} I_2^{\dagger} I_1 \ket{\psi(0)} $ and $ \bra{\psi(0)} I_1^{\dagger} I_3 \ket{\psi(0)} =\bra{\psi(0)} I_3^{\dagger} I_1 \ket{\psi(0)},$ so we have
\begin{align}\label{eq:phiphi}
    \braket{\phi}{\phi} &= \bra{\psi(0)} I_1^{\dagger} I_1 \ket{\psi(0)}+ 4B^2 \bra{\psi(0)} I_2^{\dagger} I_2 \ket{\psi(0)}  - 6B^2 \bra{\psi(0)} I_1^{\dagger} I_3 \ket{\psi(0)} +\mathcal{O}(B^3).
\end{align}
From our work above, we have $J(\rho) \leq 4 \braket{\phi}{\phi}$. Also, because $J(\rho)$ is an even function of $\omega$, $\int_{-\infty}^{\infty} J(\rho) d \omega = 2 \int_{0}^{\infty} J(\rho) d \omega := 2 J_{\text{tot}}.$ Thus,
\begin{align}
    J_{\text{tot}} &= \frac{1}{2} \int_{-\infty}^{\infty} J(\rho) d\omega, \\
    &\leq \frac{1}{2} \int_{-\infty}^{\infty} 4 \braket{\phi}{\phi} d\omega, \\
    &= 2 \int_{-\infty}^{\infty} \braket{\phi}{\phi} d\omega.
\end{align}
Thus, our task has become integrating Eq. \ref{eq:phiphi} over all frequencies. Let's start with the first term, which we write out explicitly because the same technique will be applied to the other terms. We have

\begin{align}
    &2 \int_{-\infty}^{\infty} \bra{\psi(0)} I_1^{\dagger} I_1 \ket{\psi(0)}  d\omega\\
    &= 2 \int_{-\infty}^{\infty} \bra{\psi(0)} \left(\int_{0}^{t'} d\tau' \cos{(\omega \tau')} Z_{I}(\tau')\right) \left(\int_{0}^{t} d\tau \cos{(\omega \tau)} Z_{I}(\tau)\right) \ket{\psi(0)}
    d\omega ,\\
    &= 2 \int_{-\infty}^{\infty} \bra{\psi(0)} \left(\int_{0}^{t'} d\tau' \int_{0}^{t} d\tau \cos{(\omega \tau')} \cos{(\omega \tau)} Z_{I}(\tau')  Z_{I}(\tau)\right) \ket{\psi(0)}
    d\omega .\\
\end{align}
Now, let us note the following useful fact,
\begin{align}
    2\pi \delta(\tau'-\tau) = \int_{-\infty}^{\infty} e^{i\omega(\tau'-\tau)} d \omega.
\end{align}
With this in mind, we can write
\begin{align}
    &2 \int_{-\infty}^{\infty} \bra{\psi(0)} I_1^{\dagger} I_1 \ket{\psi(0)}  d\omega\\
    &= 2 \int_{-\infty}^{\infty} \bra{\psi(0)} \left(\int_{0}^{t'} d\tau' \int_{0}^{t} d\tau \left(\frac{ \cos{\omega(\tau' - \tau)} + \cos{\omega(\tau' + \tau)}}{2} \right)Z_{I}(\tau')  Z_{I}(\tau)\right) \ket{\psi(0)}
    d\omega\\
    &= \frac{1}{2} \int_{-\infty}^{\infty} \bra{\psi(0)} \left(\int_{0}^{t'} d\tau' \int_{0}^{t} d\tau \left(e^{i\omega (\tau' -\tau)}+e^{-i\omega (\tau' -\tau)}+e^{i\omega (\tau' +\tau)}+e^{-i\omega (\tau' +\tau)}\right)Z_{I}(\tau')  Z_{I}(\tau)\right) \ket{\psi(0)}
    d\omega\\
    &= \pi \bra{\psi(0)} \left(\int_{0}^{t'} d\tau' \int_{0}^{t} d\tau \left[ \delta(\tau'-\tau) +\delta(-(\tau' -\tau))+ \delta(\tau'+\tau) +\delta(-(\tau'+\tau))\right]Z_{I}(\tau')  Z_{I}(\tau)\right) \ket{\psi(0)}\\
    &=\pi \bra{\psi(0)} \left(\int_{0}^{t'} d\tau' \int_{0}^{t} d\tau 2 \delta(\tau'-\tau)  Z_{I}(\tau')  Z_{I}(\tau)\right) \ket{\psi(0)}\\
    &=2\pi \int_{0}^{t'} d\tau' \bra{\psi(0)} Z_{I}(\tau')  Z_{I}(\tau') \ket{\psi(0)}\\
    &=2\pi \int_{0}^{t'} d\tau' \braket{\psi(0)}{\psi(0)}\\
    &=2\pi \int_{0}^{t'} d\tau'\\
    &=2\pi t',
\end{align}
where we have used the facts that $\delta(-x)=\delta(x)$, the delta functions $\delta(\tau' +\tau)  =0 $ for the range over which we are integrating, $Z_{I}(\tau')  Z_{I}(\tau') = \mathbb{I}$, and $\braket{\psi(0)}{\psi(0)}=1.$ Next, we turn to the second term in Eq. \ref{eq:phiphi}. We have
\begin{align}
    &8B^2 \int_{-\infty}^{\infty} \bra{\psi(0)} I_2^{\dagger} I_2 \ket{\psi(0)} d \omega\\
    &= 8B^2 \int_{-\infty}^{\infty} \bra{\psi(0)}\left(\int_0^{t'} d\tau'_2 \int_0^{\tau'_2} d\tau'_1 \cos(\omega \tau'_2)\cos(\omega \tau'_1)Z_I(\tau'_2) Z_I(\tau'_1)\right)\\
    & \times \left(\int_0^t d\tau_2 \int_0^{\tau_2} d\tau_1 \cos(\omega \tau_2)\cos(\omega \tau_1)Z_I(\tau_2) Z_I(\tau_1) \right) \ket{\psi(0)} d \omega,\\
    &= 8B^2 \int_{-\infty}^{\infty} \bra{\psi(0)}\left(\int_0^{t'} d\tau'_2 \int_0^{\tau'_2} d\tau'_1 \int_0^t d\tau_2 \int_0^{\tau_2} d\tau_1 \prod_i \cos{(\omega \tau_{i})} Z_{I}(\tau_{i}) \right) \ket{\psi(0)} d \omega,\\
\end{align}
where we have abused notation in attempt to compactly express the product of cosines and $Z_I$'s. As above, the integral over all frequencies kills one of the time integrals, leaving three. Finally, bounding the expectation value of the product of $Z_I$'s from above by 1, we have 
\begin{align}
    8B^2 \int_{-\infty}^{\infty} \bra{\psi(0)} I_2^{\dagger} I_2 \ket{\psi(0)} d \omega
    &\leq 16\pi B^2 t^3.
\end{align}
Similarly, for the last term we have
\begin{align}
    12 B^2 \int_{-\infty}^{\infty} \bra{\psi(0)} I_1^{\dagger} I_3 \ket{\psi(0)} d \omega \leq 24\pi B^2 t^3.
\end{align}

Together, this yields an upper bound on the IQFI of
\begin{align}
    J_{\text{tot}} \leq 2\pi t + 40 \pi B^2 t^3 + \mathcal{O}(B^3)
\end{align}
\subsection{Derivation of the IQFI under the Rotating Wave Approximation}\label{sec:RWA}
For the protocol governed by the Hamiltonian 
\begin{align}
    H(t) = \mu B \cos{\omega t} Z + \hbar g X = \begin{pmatrix} \mu B\cos{\omega t}  & \hbar g \\
    \hbar g & -\mu B\cos{\omega t}
    \end{pmatrix},
\end{align}
one can ask what the behavior of the system is near resonance ($\omega \sim 2g$). In this regime, we can apply the rotating wave approximation (RWA) to the system and still capture the dynamics. Transforming into the interaction frame with respect to $\hbar gX$ yields the RWA Hamiltonian given as
\begin{align}
    H_{\text{RWA}} &= \frac{1}{2}\begin{pmatrix} \mu B & \hbar (2g-\omega) \\
    \hbar(2g-\omega) & -\mu B
    \end{pmatrix},
\end{align}
and the time-evolution operator is $U_{\text{RWA}}(t) = \exp{\left(-\frac{i}{\hbar} H_{\text{RWA}} t \right)}$. We take $\mu=\hbar=1$ to simplify the expressions, then add the pre-factors in at the end of the calculation to restore dimensional consistency. Doing so gives a final state of the form
\begin{align}
    \ket{\psi(t)}_{\text{RWA}} &= U_{\text{RWA}}(t)\ket{+}, \\
    &= \left(
\begin{array}{c}
 \frac{\cos \left(\frac{1}{2} t \sqrt{B^2+(\omega -2 g)^2}\right)-\frac{i (B+2 g-\omega ) \sin \left(\frac{1}{2} t \sqrt{B^2+(\omega -2 g)^2}\right)}{\sqrt{B^2+(\omega -2 g)^2}}}{\sqrt{2}} \\
 \frac{\cos \left(\frac{1}{2} t \sqrt{B^2+(\omega -2 g)^2}\right)+\frac{i (B-2 g+\omega ) \sin \left(\frac{1}{2} t \sqrt{B^2+(\omega -2 g)^2}\right)}{\sqrt{B^2+(\omega -2 g)^2}}}{\sqrt{2}} \\
\end{array}
\right),\\
&:= a\ket{0} + b\ket{1},
\end{align}
where we have identified
\begin{align}
    a &= \frac{\cos \left(\frac{1}{2} t \sqrt{B^2+(\omega -2 g)^2}\right)-\frac{i (B+2 g-\omega ) \sin \left(\frac{1}{2} t \sqrt{B^2+(\omega -2 g)^2}\right)}{\sqrt{B^2+(\omega -2 g)^2}}}{\sqrt{2}}, \\
    b &= \frac{\cos \left(\frac{1}{2} t \sqrt{B^2+(\omega -2 g)^2}\right)+\frac{i (B-2 g+\omega ) \sin \left(\frac{1}{2} t \sqrt{B^2+(\omega -2 g)^2}\right)}{\sqrt{B^2+(\omega -2 g)^2}}}{\sqrt{2}},
\end{align}
for simplicity. Further defining $\partial_B [a] := c$ and $\partial_B [b] := d$ allows the QFI of estimating $B$ from this final state, $J(\omega)=4 \braket{\partial_B \psi(t)}{\psi (t)}^2 + 4\braket{\partial_B \psi(t)}{\partial_B \psi(t)}$, to be expressed as 
\begin{align}
    J(\omega) &= 4\left[ (c^*a +d^*b)^2 + c^* c + d^*d\right],\\
   &= \frac{B^2 t^2}{B^2 + (\omega-2g)^2} + \frac{4\sin^2{\left(\frac{t\sqrt{B^2+ (\omega-2g)^2}}{2}\right)}}{(B^2 +(\omega-2g)^2)^2}-\frac{t^2 B^2 (\omega -2g)^2}{(B^2+(\omega-2g)^2)^2}\\
    &+\frac{2B^2t(\omega-2g)^2 \sin{\left[t \sqrt{B^2+(\omega-2g)^2}\right]}}{(B^2+(\omega-2g)^2)^{5/2}}-\frac{B^2(\omega-2g)^2 \sin^2{\left[t \sqrt{B^2+(\omega-2g)^2}\right]}}{(B^2+(\omega-2g)^2)^3}.
\end{align}
Now, we can attempt to find the anti-derivatives of each term. Denoting the IQFI as $K(\omega)$, we find
\begin{align}
    K(\omega) &= \frac{1}{2}Bt^2 \left(\tan^{-1}\left(\frac{\omega-2g}{B}\right)+ \frac{B(\omega - 2g)}{(B^2+(\omega-2g)^2)^2}\right) \\
   &+ \int d\omega \left[\frac{4\sin^2{\left(\frac{t\sqrt{B^2+ (\omega-2g)^2}}{2}\right)}}{(B^2 +(\omega-2g)^2)^2} +\frac{2B^2t(\omega-2g)^2 \sin{\left[t \sqrt{B^2+(\omega-2g)^2}\right]}}{(B^2+(\omega-2g)^2)^{5/2}}-\frac{B^2(\omega-2g)^2 \sin^2{\left[t \sqrt{B^2+(\omega-2g)^2}\right]}}{(B^2+(\omega-2g)^2)^3}\right],\\
   &\leq t^2 \left(\frac{B}{2}\tan^{-1}\left(\frac{\omega-2g}{B}\right)+ \frac{B^2(\omega - 2g)}{2(B^2+(\omega-2g)^2)^2}\right) \\
   &+ \int d\omega \left[\frac{4}{(B^2 +(\omega-2g)^2)^2} +\frac{2B^2t(\omega-2g)^2 }{(B^2+(\omega-2g)^2)^{5/2}}-\frac{B^2(\omega-2g)^2 }{(B^2+(\omega-2g)^2)^3}\right],\\
   K(\omega) &=  t^2 \left(\frac{B}{2}\tan^{-1}\left(\frac{\omega-2g}{B}\right)+ \frac{B^2(\omega - 2g)}{2(B^2+(\omega-2g)^2)^2}\right) + \mathcal{I}(t),
\end{align}
where we have let $\mathcal{I}(t)$ represent the integral. When integrated, the result is at most linear in $t$. Focusing on the first term, which dominates for $T \gg 1$, we can evaluate the anti-derivative to recover the expression in the main text. The RWA holds in a frequency band around the resonant peak, which we take to be $\omega = g$ to $\omega = 3g.$ Evaluating the anti-derivative over this band, we obtain 
\begin{align}
   K(T) \gtrsim  B \mu  T^2 \left(\frac{B g \mu }{B^2 \mu ^2+g^2}+\tan ^{-1}\left(\frac{g}{B \mu }\right)\right).
\end{align}
Restoring dimensional consistency, we obtain the expression from the main text
\begin{align}
    K(T) \gtrsim   \zeta^2T^2 \left(\frac{ g }{1+ \frac{g^2}{\zeta^2 B^2}}+\zeta B\tan ^{-1}\left(\frac{g}{\zeta B}\right)\right).
\end{align}

\subsection{Bound on IQFI of Continuous Protocols}\label{sec:gen_proof}
As in the main text, consider a sensing protocol defined by a time-dependent control $G(T)$. In particular, the full Hamiltonian we are considering is 
\begin{equation}
    H(T) = \hbar G(T) + \mu B\cos{(\omega T)}Z.
\end{equation}
We can Trotterize the evolution into a discrete sequence \cite{poulin2011quantum}, which will look like those considered in the proof above, with some step size $\delta t$. The evolution will be
\begin{equation}\label{eq:trotter}
    U'(T) = \mathcal{T}e^{i\int_{(N-1)\delta t}^{N\delta t} dt  G(t)}e^{i\zeta B\Theta_NZ} \dotsm \mathcal{T}e^{i\int_0^{\delta t} dt G(t)}e^{i\zeta B\Theta_1Z} + O(\delta t^2)
\end{equation}
with $\Theta_k = [\sin{(\omega k\delta t)} - \sin{(\omega (k-1)\delta t)}]/(\hbar\omega)$, $N\delta t = T$, and where $\mathcal{T}$ denotes the time-ordering operator, which is necessary because in general $G(t)$ will not commute with itself at all times. The number of pulses, $N$, in the discrete protocol described in Eq.~\eqref{eq:trotter} gives, to zeroth order,
\begin{equation}\label{eq:zeroth}
    \int d\omega J_c(\omega) \leq \frac{2\pi \zeta^2 T^2}{\delta t},
\end{equation}
for all $\delta t$. There are however error terms from the Trotter expansion that we need to propagate through the IQFI - this is what we will do now. By linearity of the derivative operator we can also constrain the derivative of the Trotterized evolution to be close to the derivative of the actual evolution. In particular, we have that
\begin{equation}
    U(T) = U'(T) + \hat{\epsilon}'(T)
\end{equation}
\begin{equation}
    \partial_B U(T) = \partial_B U'(T) + \hat{\epsilon}''(T)
\end{equation}
where $\hat{\epsilon}''(T) = \partial_B\hat{\epsilon}'(T)$ is the B derivative of the error term in the Trotter expansion. Then we see from Eq.~\eqref{eq:pure_qfi_bound},
\begin{eqnarray}
    J_c(\omega) \leq J_d(\omega) + 4(\bramatket{+}{\hat{\epsilon}''^{\dagger}(T)\partial_B U'(T) + \partial_B U'^{\dagger}(T)\hat{\epsilon}''(T)}{+} + \bramatket{+}{\hat{\epsilon}''(T)^{\dagger}\hat{\epsilon}''(T)}{+})\\ = J_d(\omega) + 8Re(\bramatket{+}{\hat{\epsilon}''(T)^{\dagger}\partial_B U'(T)}{+}) + 4\bramatket{+}{\hat{\epsilon}''(T)^{\dagger}}{\hat{\epsilon}''(T)|+}).\label{eq:naive_inequality}
\end{eqnarray}
We will analyze the integrals of the error terms piecewise, first up to a frequency $\Omega > 0$. The derivative of the first order Trotterized evolution defined above is
\begin{equation}\label{eq:trotter_derivative}
    \partial_B U'(T) = -i\zeta\sum_{j=0}^{N-1} \Theta_jP_0U_0...P_jZU_j...P_{N-1}U_{N-1}.
\end{equation}
Noting that $|\Theta_k| \leq 2T/N$ for a uniformly spaced pulse sequence, this gives
\begin{eqnarray}
    ||\int_0^{\Omega} \hat{\epsilon}''(T)(-\frac{i\mu}{\hbar})\sum_{j=0}^{N-1} \Theta_jP_0U_0...P_jU_j...P_{n-1}U_{n-1}||\\ \leq \frac{\mu}{\hbar}{\rm max_{\omega<\Omega}}||\hat{\epsilon}''(T)||\int_0^{\Omega}\sum_{j=0}^{N-1} |\Theta_j|||P_0U_0...P_jU_j...P_{n-1}U_{n-1}||\\ \leq \frac{\mu}{\hbar}{\rm max_{\omega<\Omega}}||\hat{\epsilon}''(T)||\int_0^{\Omega}\sum_{j=0}^{N-1} 2T/N \\= \frac{2\mu}{\hbar}{\rm max_{\omega<\Omega}}||\hat{\epsilon}''(T)||\Omega T,
\end{eqnarray}
Similarly,
\begin{eqnarray}
4\int_0^\Omega||\hat{\epsilon}''(T)^{\dagger}\hat{\epsilon}''(T)|| \leq 4{\rm max_{\omega<\Omega}}||\hat{\epsilon}''(T)||^2\Omega
\end{eqnarray}

where the maximum is over $\omega$ on which the error term implicitly depends. Then we have, for the last two terms in Eq.~\eqref{eq:naive_inequality}
\begin{equation}\label{eq:up_to_omega}
    8||\int_0^{\infty} Re(\hat{\epsilon}''^{\dagger}(T)\partial_B U'(T))|| \leq \frac{16\mu}{\hbar}\Omega T{\rm max_{\omega<\Omega}}||\hat{\epsilon}''(T)|| + ||\int_\Omega^{\infty} 8Re(\hat{\epsilon}''^{\dagger}(T)\partial_B U'(T))||,
\end{equation}
and
\begin{eqnarray}\label{eq:up_to_omega2}
    4\int_0^\infty||\hat{\epsilon}''(T)^{\dagger}\hat{\epsilon}''(T)|| \leq 4{\rm max_{\omega<\Omega}}||\hat{\epsilon}''(T)||^2\Omega +  4\int_\Omega^{\infty}||\hat{\epsilon}''(T)^{\dagger}\hat{\epsilon}''(T)||.
\end{eqnarray}
$\hat{\epsilon}''(T)$ is at worst proportional to $T$, since the error in a time step $\delta t$ is independent of $T$ (it's proportional to $||[\hbar G(t), \mu B\cos{(\omega t)}Z]||\delta t^2$), and there are $N=T/\delta t$ times steps. Furthermore
\begin{equation}
    ||\partial_BU'(T)|| = ||\frac{-i\mu}{\hbar}\sum_j \Theta_jP_1U_1...P_jU_j...P_nU_n|| \leq \mu N/(\hbar \omega) = \mu T/(\delta t\hbar\omega),
\end{equation}
so we need only understand $||\hat{\epsilon}''(T)||$ at high frequency, where $||\cdot||$ is the spectral norm. To this end, consider the error in time step $\delta t$ given in \cite{Huyghebaert1990} as $U'(t_k, t_{k-1})F(t_k, t_{k-1})$, where $U'(a,b)$ is the Trotterized evolution from time $a$ to time $b$ and 
\begin{eqnarray}
    F(t_k, t_{k-1}) = \int_{t_{k-1}}^{t_k} C(v, t_{k-1})F(v, t_{k-1})dv\\
    C(t_k, t_{k-1}) = \exp{(i\int_{t_{k-1}}^{t_k}G(s)ds)}^{\dagger}\int_{t_{k-1}}^{t_k}du\  \exp{(i\int_{t_{k-1}}^{u}A(s)ds)}^{\dagger}\\
    \times [A(u), \hbar G(t_k)]\exp{(-i\int_{t_{k-1}}^{u}A(s)ds)}\exp{(-i\int_{t_{k-1}}^{t_k}G(s)ds)},
\end{eqnarray}
where $A(s) = \frac{\mu B}{\hbar}\cos{(\omega s)}Z$. Thus, to lowest order in $1/\omega$,
\begin{equation}
    \exp{(-i\int_{t_{k-1}}^{u}A(s)ds)} = \exp{(iBZ(\cos{(\omega u)} - \cos{(\omega t_{k-1})})/\omega)} \approx I - i\frac{\mu B}{\hbar}Z(\cos{(\omega u)} - \cos{(\omega t_{k-1})})/\omega
\end{equation}
so that the relevant integrals over $u$ are 
\begin{eqnarray}
    \int_{t_{k-1}}^{t_k}du \cos{(\omega u)} = (\sin{(\omega t_k)} - \sin{(\omega t_{k-1})})/\omega\\
    \int_{t_{k-1}}^{t_k}du \cos{(\omega u)}(\cos{(\omega u)} - \cos{(\omega t_{k-1})}) \\= (2\delta t\omega + \sin{(2t_{k-1}\omega}) - 4\cos{(t_{k-1}\omega)}\sin{((t_{k-1} + \delta t)\omega)} + \sin{(2(t_{k-1} + \delta t)\omega)})/(4\omega)
\end{eqnarray}
Plugging these in we find
\begin{eqnarray}\label{eq:error_term}
    C(t_k, t_{k-1}) = \frac{\mu B}{\hbar}((\sin{(\omega t_k)} - \sin{(\omega t_{k-1})})/\omega)(\exp{(i\int_{t_{k-1}}^{t_k}\hbar G(s)ds)}^{\dagger}[Z, \hbar G(t_k)]\exp{(-i\int_{t_{k-1}}^{t_k}\hbar G(s)ds)}) \\
    + \frac{i2\zeta^2 B^2}{4\omega^2}((2\delta t\omega + \sin{(2t_{k-1}\omega}) - 4\cos{(t_{k-1}\omega)}\sin{((t_{k-1} + \delta t)\omega)} + \sin{(2(t_{k-1} + \delta t)\omega)})\\\times(\exp{(i\int_{t_{k-1}}^{t_k}\hbar G(s)ds)}^{\dagger}Z[Z, \hbar G(t_k)]\exp{(-i\int_{t_{k-1}}^{t_k}\hbar G(s)ds)})\\ + (\exp{(i\int_{t_{k-1}}^{t_k}\hbar G(s)ds)}^{\dagger}[Z, \hbar G(t_k)]Z \exp{(-i\int_{t_{k-1}}^{t_k}\hbar G(s)ds)}))))),
\end{eqnarray}
where we have only kept the lowest order correction terms. By inspection, we see that there is no way for a $T$ dependence to enter for higher order terms, and moreover we see that for $\omega$ large enough ($\Omega >> \mu B/\hbar$) the terms are both integrable over $\omega$ and arbitrarily small. We are interested in $N$ times the B-derivative of this error, where the extra factor of $N$ is because the total error accumulates at worst as $N$ times the step-wise error. But now we are done, since this shows that for all $\epsilon > 0$ there is an $\Omega(||G||, B, \delta t)$ such that in total we have a bound of
\begin{equation}
    \int d\omega J_c(\omega) \leq \frac{2\pi \zeta^2 T^2}{\delta t} + (c(\hbar||G||, \mu B, \delta t)  + \epsilon(\hbar||G||, \mu B, \delta t))T^2 = \frac{2\pi \zeta^2 T^2}{\delta t} + \alpha(\hbar||G||, \mu B, \delta t)T^2,
\end{equation}
where we've denoted two parts in the coefficient of the second term - one coming from the lowest order contribution to the right tail of the QFI, and one coming from the inaccuracy in this approximation. So in total we see that $\int d\omega K(T) \in \mathcal{O}(T^2)$. We see furthermore that this argument generalizes to arbitrary entangled probes. The $B-$derivative of the time evolution is maximized on entangled inputs and contributes a factor of $n$, the number of qubits. To see this, note Eq.~\eqref{eq:trotter_derivative} for $n$ qubits is instead

\begin{equation}
    \partial_B U'(T) = -i\zeta\sum_{j=0}^{N-1}\sum_{k=0}^{k=n-1} \Theta_jP_0U_0...P_jZ^{(k)}U_j...P_{N-1}U_{N-1},
\end{equation}
where the new index $(k)$ denotes the $Z$ operation on the $k^{th}$ qubit, and the other operators are generalized to their natural multi-qubit counterparts. Similarly, the Trotter error has a factor of $n$ because it is proportional to the sum of $Z^{(k)}$, and so the $B-$derivative of the Trotter error also is proportional to $n$. Because both terms in the QFI are quadratic in $\ket{\phi}$ we find a factor of $n^2$ as in the GHZ state example, giving a total bound of $\int d\omega K(T) \in \mathcal{O}(n^2T^2)$ for $n$ qubits.

\bibliography{nofreeqfi.bib}

\end{document}